\documentclass[12pt]{iopart}

\usepackage{amssymb}
\usepackage{amsfonts}
\usepackage{amstext}
\usepackage{graphicx}
\usepackage{amscd}

\newcommand{\ket}[1]{\vert{#1}\rangle}

\begin{document}

\title[Discussion of the adiabatic hypothesis in control schemes using EP]{Discussion of the adiabatic hypothesis in control schemes using exceptional points}
\author{A Leclerc$^{1,2}$, G Jolicard$^1$ and J P Killingbeck$^3$}

\address{$^1$ Institut UTINAM, CNRS UMR 6213, Universit\'e de Franche-Comt\'e, Observatoire de Besan\c con,
41 bis Avenue de l'Observatoire, BP 1615, 25010 Besan\c con cedex, France\\
$^2$ Chemistry Department, Queen's University, Kingston, Ontario K7L 3N6, Canada \\
$^3$ Centre for Mathematics, University of Hull, Hull HU6 7RX, UK}

\ead{Arnaud.Leclerc@utinam.cnrs.fr}

\submitto{\JPB}

\begin{abstract}
We present calculations for the action of laser pulses on vibrational transfer within the H$_2^+$ and Na$_2$ molecules
in the presence of dissipation due to photodissociation of the molecule. 
The laser fields perform closed loops surrounding exceptional points in the laser parameter plane of intensity and wavelength. 
In principle the process should produce controlled vibrational transfers due to an adiabatic flip of the dressed eigenstates. 
We directly solve the Schr\"odinger equation with the complete time-dependent field instead of using the adiabatic Floquet 
formalism which initially suggested the design of the laser pulses. Results given by wavepacket propagations disagree with 
predictions obtained using the adiabatic hypothesis. Thus we show that there are large non-adiabatic exchanges and that the 
dissipative character of the dynamics renders the adiabatic flip very difficult to obtain.
Using much longer durations than expected from previous studies, the adiabatic flip is only obtained for the Na$_2$ molecule and with strong dissociation.
\end{abstract}

%
%

\section{Introduction}

An effective quantum mechanical treatment of processes such as photodissociation can be obtained by using non-Hermitian Hamiltonians \cite{NHQM}
which depend on two or more parameters and so have singularities named exceptional points (EP)
\cite{kato}. An EP is a coalescence of two eigenvalues and two eigenstates for a certain set of parameters for which the 
Hamiltonian becomes non-diagonalizable. EP processes can induce dramatic physical effects, which are well summarized in 
\cite{heiss2012} and references therein. The EP phenomenon is also referred to as self-orthogonality because of a coalescence 
of eigenvectors which does not happen for Hermitian Hamiltonian degeneracies \cite{chap_EP_moiseyev}. Among the many papers 
on this subject, we can cite some showing the manifestations of EP: in microwave cavities \cite{dembowski2001,dietz}, 
for $\mathcal{ PT}$ symmetry in optics \cite{longhi2010} and waveguides \cite{klaiman2008}, in atomic and molecular physics
\cite{lefebvre,atabek2,lefebvre2,lefebvre3,atabek3}, in laser physics \cite{liertzer2012} and in quantum phase transitions \cite{heiss2005}.
An adiabatic separation is assumed between the two Floquet resonances related to the EP and all other states.
Here we should mention that similar assumptions are present in studies on interference stabilisation of ionisation in Rydberg atoms, or stabilisation 
of dissociation in molecules \cite{fedorov1,fedorov2}. In these two cases, a strong laser field provides ionisation or dissociation from two dissipative levels and their reconstruction owing to Raman-type transitions. This interference stabilisation reduces significantly the ionisation or the dissociation yields. Coming back to the EPs, 
an interesting effect is the adiabatic flip: the two eigenstates concerned in the EP interchange when the parameter trajectory 
encircles the EP. This feature could be particularly useful in the context of laser control problems. 
The point of our study is to show that the presence of an EP, which should produce the interchange of the two eigenstates concerned in the EP, is simultaneously unfavourable to the adiabaticity of the process. Then longer durations than previously calculated are necessary to obtain controlled transitions, which are not so efficient.

For atomic and molecular systems in the presence of an intense laser (periodic) field, use can be made of the Floquet Hamiltonian of the form \cite{shirley}
\begin{equation*}
\tilde{H}_F ({q},\theta) = H_0(q) - \vec \mu . \vec E_0 \cos (\theta) - i \hbar \omega \frac{\partial}{\partial \theta},
\end{equation*}
where $H_0$ stands for the field-free molecular hamiltonian and the second term is the electric dipole coupling, if we 
adopt the length-gauge and the long wavelength approximation. 
$\omega$ is the optical frequency and $\theta = \omega t \in [0,2\pi] $. 
Floquet eigenstates 
(such that $\tilde{H}_F \ket{\tilde{\Phi}} = \tilde{E}_{\Phi} \ket{\tilde{\Phi} }$) are then related to solutions of the Schr\"odinger 
equation $\psi(q,\theta)$ by the equation $\psi(q,\theta) = e^{-iE_{\Phi} \theta / ( \hbar \omega) } \tilde{\Phi} (q,\theta)$. 
The adiabatic Floquet Hamiltonian is a generalization of $\tilde{H}_F$ when the laser
parameters are varying~\cite{reviewguerin}:
\begin{equation}
 H_F ({q},t,\theta) = H_0(q) - \vec \mu . \vec E_0(t) \cos (\omega (t) t) - i \hbar \omega_{\text{eff}} (t) \frac{\partial}{\partial \theta}.
 \label{Hfadia}
\end{equation}
Here $\theta $ is the rapid phase associated with the optical oscillations and $t$ stands for the slow time scale corresponding to the variations of the field parameters 
(field amplitude $E_0(t)$ and effective angular frequency $\omega_{\text{eff}}(t)$, or equivalently field intensity $I \propto E_0^2$ and wavelength $\lambda$).
The effective frequency and the real frequency are related by 
\begin{equation}
\omega_{\text{eff}} = \frac{d}{dt} \left( \omega (t) t \right). 
\label{omegaeff}
\end{equation}
The Floquet resonances (such that $H_F(t) \ket{\Phi(t)} = E_{\Phi}(t) \ket{\Phi (t)}$) are then obtained by rendering the Hamiltonian non-hermitian, using complex absorbing potentials \cite{CAP2} or a complex scaling transformation \cite{moiseyev1998} so as to obtain an operator which can describe photodissociation phenomena.
For such a Hamiltonian EPs exist and connect two instantaneous Floquet resonances, which are themselves connected to given vibrational eigenstates of the field-free molecule.
Adiabatic flips has recently been predicted for the H$_2^+$ ion \cite{lefebvre,atabek3,lefebvre3} and for cold Na$_2$ molecules 
\cite{lefebvre2,atabek2}, occurring between two vibrational states $v$ and $v\pm 1$ or $v\pm 2$. A chirped pulse is expected to 
cause a selective transfer of the vibrational population, permitting for example the progressive vibrational cooling of the 
Na$_2$ molecules which has not dissociated under the action of the pulse. 
For this, an adiabatic assumption is made concerning the instantaneous dressed Floquet states $\ket{\Phi(t)}$ (eigenvectors 
of $H_F$ with $E_0(t)$ and $\omega(t)$ fixed). Slow variations of the intensity and of the frequency of the field are assumed, 
together with the presence of a gap between the followed eigenvalue and the other eigenvalues. 
In this context, the dissociation probability is estimated using the adiabatic formula
\begin{equation}
 P_{\text{diss}}(t) =1- \exp \left[ - \hbar^{-1} \int_0^{t} \Gamma (t') dt' \right]
 \label{pdissfloqadiab}
\end{equation}
where $\Gamma/2$ is the imaginary part (width) of the quasienergy $E_{\Phi}$ calculated with the field parameters at time $t'$.

These works using adiabaticity have been partially contradicted by recent papers which have pointed out an asymmetry of the adiabatic flip
\cite{uzdin,gilary2,leclercviennot,berrycycling}. In these papers the dynamics was described in terms of the instantaneous eigenstates 
but the calculation also took into acount non-adiabatic exchanges between the two coalescing states within a two-dimensional subspace. 
The adiabatic hypothesis is then well-satisfied only following the less dissipative eigenstate, in accordance with known adiabatic theorems 
for dissipative Hamiltonians \cite{nenciurasche}.

Here we treat the problem without any adiabatic approximation, using standard wavepacket propagation and an explicitly time-dependent laser pulse in the hamiltonian. 
We try to check numerically the flip effect due to an EP for the two above diatomic systems for which pulses have been already designed in the literature. 
Thus we do not use the adiabatic Floquet Hamiltonian of (\ref{Hfadia})
but we propagate the wavefunctions directly from the Schr\"odinger equation
using the Hamiltonian $H_0 - \vec \mu . \vec E_0(t) \cos (\omega (t) t)$. 
The paper is organized as follows.
In section \ref{modelsec} we present the model used to describe the photodissociative dynamics of H$_2^+$ or Na$_2$. Section \ref{floquetpredictions} recalls the population transfers as predicted by the adiabatic Floquet formalism. Then the contradictory results given by wavepacket propagations are presented in section \ref{WPresults}. 
Finally we use longer pulse durations to obtain a partial and asymmetric state interchange, only for the Na$_2$ case.  


\section{Photodissociation model \label{modelsec}}

Let us recall briefly the model used to describe the photodissociation of a diatomic molecule. 
We use the same numerical data as in \cite{lefebvre,atabek3,lefebvre3,lefebvre2,atabek2}.
We consider a one-dimensional model with the internuclear distance $R$. The laser light is assumed to be linearly polarized. 
In the framework of the Born-Oppenheimer approximation, we will only consider two electronic states labelled $\ket{1}$ and $\ket{2}$ so that the wavefunction is 
\begin{equation}
 \Psi(R,t) = \chi_1(R,t) \ket{1} + \chi_2(R,t)  \ket{2}.
\end{equation}
In our calculations, the dynamics of the nuclear wavefunction is then governed by the Schr\"odinger equation

  
\begin{eqnarray}
i\hbar \frac{\partial}{\partial t}
\left(
\begin{array}{l}
\chi_1(R,t) \\ 
\chi_2(R,t)
       \end{array}
\right)
& =  \left[ T_N + \left( 
\begin{array}{ll}
\epsilon_1 (R) & 0 \\ 
0 & \epsilon_2 (R)
 \end{array}
 \right)  \right. \nonumber \\ 
 & 
 \left.
-\vec{\mu}_{12} (R) \cdot \vec{E}(t)
\left( 
\begin{array}{ll}
0 & 1 \\ 
1 & 0
 \end{array}
 \right)
\right]
\left(
\begin{array}{l}
\chi_1(R,t) \\ 
\chi_2(R,t)
       \end{array}
\right).
\label{EP_contexte}
\end{eqnarray}

$T_N$ represents the kinetic energy and $\epsilon_{1/2}$ are the effective electronic potential curves. The dipole moment is such that $\mu_{12}=\mu_{21}$.
For H$_2^+$ we use the two lowest electronic states $^2\Sigma_g^+$ (labelled as state $\ket{1}$) and $^2\Sigma_u^+$ (labelled as state $\ket{2}$) 
and the dipole moment as given in \cite{bunkin}. The hydrogen ion is submitted to a laser pulse with a wavelength around $420$ nm. For Na$_2$ 
we consider two excited electronic states. We assume that the molecule is prepared in a given vibrational level of the lowest triplet state 
$^3\Sigma_u^+ (3^2S + 3^2S)$ (simply noted state $\ket{1}$) \cite{magnier}. This situation is realistic, since sodium molecules has been 
experimentally produced in this electronic state after photoassociation of cold atoms \cite{fatemi}. A laser pulse with wavelength around 
$560$ nm will allow transitions to the state $(1)^3\Pi_g (3^2S+3^2P)$ (labelled state $\ket{2}$) \cite{aymar2005,jaouadi}. 
In both systems the first electronic state supports bound states and a continuum (which is discretized for numerical analysis) while the 
second electronic state is purely 
repulsive.
In both cases the energy of the photons is sufficient to dissociate the molecules. 
The only difference of our model from that of references \cite{lefebvre,atabek3,lefebvre3,lefebvre2,atabek2} 
is the use of a polynomial complex absorbing potential at the edge of the grid instead of a complex scaling method. However, these two approaches 
are closely related and their effects are similar \cite{rom,rissmeyer,santra}. 
Convergence for varying absorbing potential has been checked. 
More precisely, the absorbing potential is defined at the edge of the radial grid by a power function 
$V_{\text{abs}} (R) = - i A \left( \frac{R-R_{\text{start}}}{R_{\text{max}}-R_{\text{start}}} \right)^{16}$, $A\in \mathbb{R}$,
and we have checked the stability of the results with respect to $A$ and $R_{\text{start}}$. 

\section{Laser pulses and adiabatic predictions \label{floquetpredictions}}

\begin{figure}[htp]
 \centering
\includegraphics[width=0.8\linewidth]{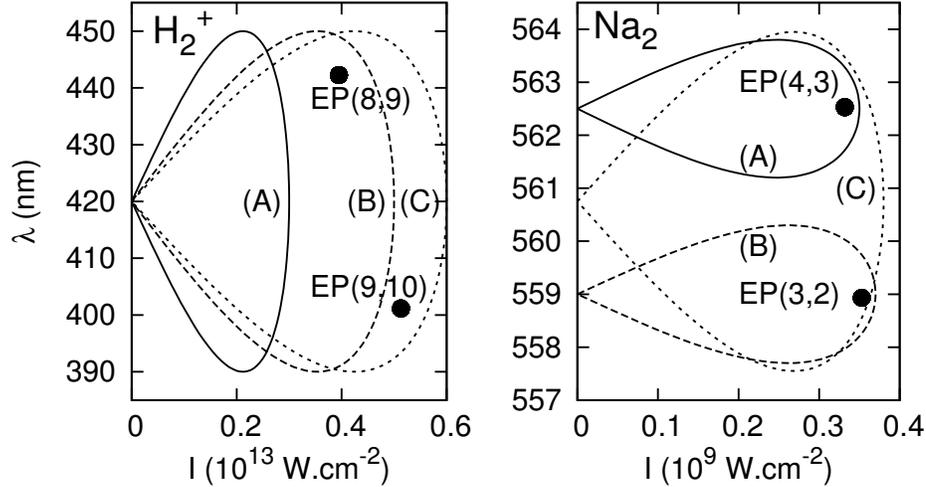}
 \caption{Paths followed in the parameter plane $(I,\lambda)$ for H$_2^+$ (left) or Na$_2$ (right). Some of them encircle EPs associated with vibrational 
 states (H$_2^+$: $v=8,9$ and $v=9,10$; Na$_2$: $v=3,4$ and $v=2,3$). Contours are followed clockwise. Numerical values for the parameters are given in tables \ref{valnumEPH2} and \ref{valnumEPNa2}. (Figure after \protect\cite{lefebvre,atabek2})}
 \label{contours_H2+}
\end{figure}

Let us sum up some results given by adiabatic Floquet theory when the aim is to obtain a selective transfer from one vibrational state to another.
R. Lefebvre, O. Atabek and their coworkers made the first detailed analysis of the EPs in the Floquet  spectrum for the two 
diatomic systems under study. For H$_2^+$ they found that the Floquet resonance associated with the vibrational state $v=8$ coalesces with the resonance 
associated with $v=9$ when $\lambda_{EP}^{8-9}=442.26$ nm and $I_{EP}^{8-9}=0.3949\times10^{13}$ W.cm$^{-2}$ \cite{lefebvre,atabek3}, among others.
Two EP are shown in figure~\ref{contours_H2+}, together with the paths proposed to obtain selective transfers of vibrational population.

The contours are defined by
\begin{equation}
 I=I_{\text{max}} \sin (\phi/2), \qquad \lambda = \lambda_0+ \delta \lambda \sin(\phi),
 \label{eqcontour}
\end{equation}
with $\phi= 2\pi t / T_{\text{tot}}$, where $T_{\text{tot}}$ is the total duration of the pulse. 
This definition is not the only one possible but has been motivated by the three following constraints:
to start and finish with null intensity, encircling the EP 
(depending on $\delta \lambda$ and $I_{\text{max}}$) 
and showing smooth variations. 
%
The dynamical process begins from a given vibrational state (for example $v=8$).
We now briefly recall some of the exchanges of population predicted using the adiabatic Floquet theory, using a pulse duration of $T_{\text{tot}} = 56$ fs $\simeq 2315$  atomic units, which was proposed as a compromise to avoid total photodissociation \cite{lefebvre}: 
\begin{itemize}
 \item begin in $v=8$, return to initial state $v=8$ using pulse (A) of figure~\ref{contours_H2+}, with a final dissociation probability of $0.75$;
 \item transfer $v=8 \rightarrow v=9$ using pulse (B), dissociation probability about $0.999$;
 \item transfer $v=9 \rightarrow v=8$ using pulse (B), dissociation about $0.90$;
 \item transfer $v=8 \rightarrow v=10$ using pulse (C).\\
\end{itemize}


For the cold sodium molecule Na$_2$, the EPs have been determined in \cite{atabek2,lefebvre2,lefebvre3}. We now recall the examples treated in \cite{atabek2}. 
Pulses have also been designed to encircle some of these points; they are given by (\ref{eqcontour}). The paths to follow are shown in figure~\ref{contours_H2+}. The sodium is less dissociative than H$_2^+$, so it is possible to choose a longer total duration of $800$ fs $\simeq 33073$ a.u. 
Here are some expected results according to the adiabatic Floquet hypothesis \cite{atabek2}: 
\begin{itemize}
 \item vibrational transfers are expected, $v\rightarrow v-1$ or $v-1\rightarrow v$ for contours (A) or (B) which encircle one of the EP;
 \item the dissociation probability given by (\ref{pdissfloqadiab}) is weak ($ P_{\text{diss}} (T_{\text{tot}})\simeq 0.20$) in the case of the inversion $v=4\rightarrow 3$; 
 \item the transfer $3\rightarrow 4$ is more dissociative, $P_{\text{diss}} (T_{\text{tot}}) \simeq 0.85$;
 \item the use of a pulse (C) which encircles both EPs implies a direct passage from $v=4$ to $v=2$ with dissociation about $0.22$.
 \end{itemize}
Within the adiabatic approximation, all the non dissociated wavepacket is always assumed to be flipped at the end of the pulse.

\section{Contradictory results given by wavepacket propagations \label{WPresults}}
 

We now report results given by using time-dependent wavepacket propagations which take into acount all non-adiabatic exchanges.
The wavepacket propagations use the constrained adiabatic trajectory method (CATM) explained in detail in \cite{CATM,CATM2,CATM3,CATM4}. 
Here the adjective 'adiabatic' has a special meaning which is different from the usual one: it simply reflects the fact that the rigorous 
wavefunction is obtained via a single global Floquet eigenstate associated with the complete duration of the pulse. Indeed, no adiabatic approximation is 
made in the usual sense.
The oscillating field is explicitly calculated as $ E(t) = E_0 (t) \cos (\omega(t) t )$ and the variations of $E_0$ and $\omega$ are given by eqs.~\ref{omegaeff} and \ref{eqcontour}. 
The CATM is implemented with a 8192 sampling for the time-dimension. 
Knowing the evolution of the wavefunction, we calculate the probability to occupy the $v^{\text{th}}$ bound state by $P_v(t)=|\langle v | \chi_1 (t)\rangle |^2$ and the dissociation probability as
\begin{equation}
 P_{\text{diss}} (t) = 1 - \sum_{\text{bound states}} P_v(t).
\end{equation}

\begin{figure}%
\centering
\includegraphics[width=0.8\linewidth]{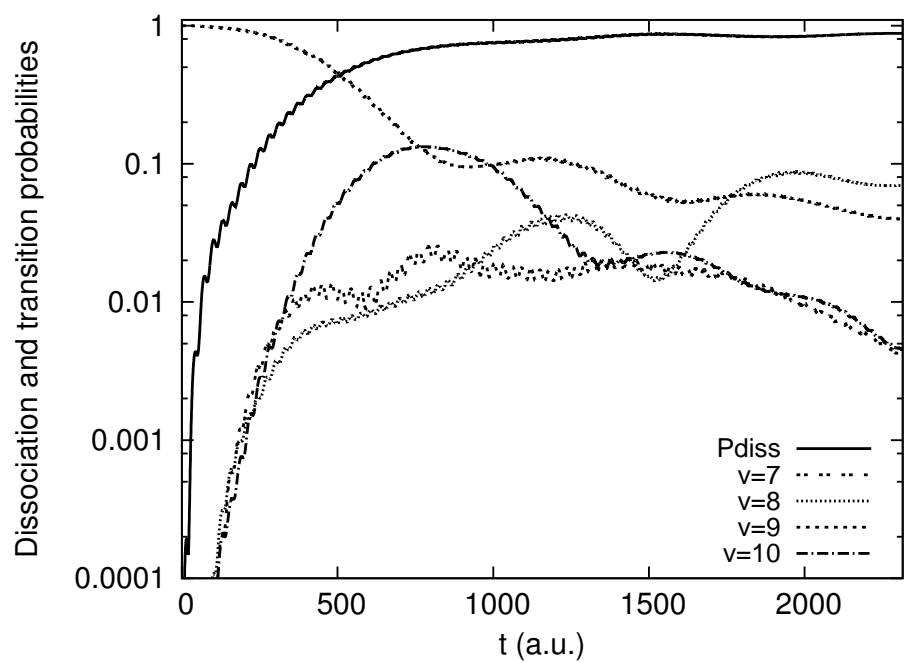}
\caption{Time evolution of the dissociation and transition probabilities, for case (d)
of table \ref{valnumEPH2} (H$_2^+$ case). 
Time is indicated in atomic units (1 a.u. $\simeq 2.42\times 10^{-17}$ s). 
}%
\label{fig:pdissvEPH2}%
\end{figure}

Table \ref{valnumEPH2} shows the numerical parameters used to calculate the wavepacket evolution of H$_2^+$. 
Table \ref{valnumEPNa2} contains equivalent information concerning Na$_2$. 
Some transition probabilities and the dissociation probability at the end of each pulse are shown in table \ref{probasEPH2_tab} for the different cases and figure~\ref{fig:pdissvEPH2} represents the time evolution of these probabilities for run (d).

\begin{table}[htp]
 \centering
\begin{tabular}{llllll}
\hline
\hline
Run & Path & $I_{\text{max}} $ ($10^{13}$W.cm$^{-2}$) & $\lambda_0$ (nm) & $\delta \lambda$ & initial $v$  \\ 
\hline
(a) & (A) & 0.3 & 420 & 30 & 8 \\
(b) & (A) & 0.3 & 420 & 30 & 9 \\
(c) & (B) & 0.5 & 420 & 30 & 8 \\
(d) & (B) & 0.5 & 420 & 30 & 9 \\
\hline
\end{tabular}
\caption{Choice of the contour in the laser parameter plane and of the initial state (cf. (\ref{eqcontour})), for H$_2^+$.}
\label{valnumEPH2}
\end{table}


\begin{table}[ht]
 \centering
\begin{tabular}{llllll}
\hline
\hline
Run & Path & $I_{\text{max}} $ ($10^9$W.cm$^{-2}$) & $\lambda_0$ (nm) & $\delta \lambda$ (nm) & initial $v$  \\ 
\hline
(a') & (A) & 0.35 & 562.5 & 1.3 & 4 \\
(b') & (A) & 0.35 & 562.5 & 1.3 & 3 \\
(c') & (B) & 0.37 & 559 & 1.3 & 3 \\
(d') & (B) & 0.37 & 559 & 1.3 & 2 \\
(e') & (C) & 0.38 & 560.75 & 3.2 & 4 \\
(f') & & 0.60 & 562.5 & 2 & 4 \\
\hline
\end{tabular}
\caption{Same as Table \ref{valnumEPH2} for Na$_2$. }
\label{valnumEPNa2}
\end{table}


 \begin{table}[htp]
 \centering
 {\small
\begin{tabular}{lllll}
\hline
\hline
 Run & (a) & (b) & (c) & (d)\\ 
\hline
 $P_{\text{diss}}$& $0.7099$ & $0.9335$ & $0.8646$ & $0.8799$ \\
 $P_0$& $3.42\times 10^{-13}$&$2.41\times 10^{-13}$ &$4.12\times 10^{-13}$&$1.45\times 10^{-12}$\\
 $P_1$& $3.59\times 10^{-11}$&$2.26\times 10^{-11}$ &$4.78\times 10^{-11}$&$1.41\times 10^{-10}$\\
 $P_2$& $1.68\times 10^{-9}$&$9.09\times 10^{-10}$& $2.46\times 10^{-9}$&$6.05\times 10^{-9}$\\
 $P_3$& $3.87\times 10^{-8}$&$1.82\times 10^{-8}$ &$9.22\times 10^{-8}$ &$1.35\times 10^{-7}$\\
 $P_4$& $3.85\times 10^{-7}$& $1.37\times 10^{-7}$&  $3.34\times 10^{-6}$ &$1.82\times 10^{-6}$\\
 $P_5$& $2.43\times 10^{-5}$& $3.65\times 10^{-6}$&  $1.42\times 10^{-5}$ &$4.86\times 10^{-5}$\\
 $P_6$& $1.14\times 10^{-4}$& $2.65\times 10^{-5}$&  $3.02\times 10^{-4}$ &$2.56\times 10^{-4}$\\
 $P_7$& $\mathbf{ 8.30\times 10^{-3}}$& $\mathbf{7.26\times 10^{-4}}$&  $\mathbf{7.25\times 10^{-3}}$ &$\mathbf{4.18\times 10^{-3}}$\\
 $P_8$& $\underline{\mathbf{0.2788}}$& $\mathbf{2.67\times 10^{-2}}$&  $\underline{\mathbf{0.1265}}$ &$\mathbf{6.97\times 10^{-2}}$\\
 $P_9$& $\mathbf{1.73\times 10^{-3}}$& $\underline{\mathbf{3.47\times 10^{-2}}}$& $3.62\times 10^{-4}$ &$\underline{\mathbf{3.99\times 10^{-2}}}$\\
 $P_{10}$& $\mathbf{8.76\times 10^{-4}}$&$\mathbf{3.80\times 10^{-3}}$ & $\mathbf{4.37\times 10^{-4}}$ &$\mathbf{4.55\times 10^{-3}}$\\
 $P_{11}$& $3.45\times 10^{-5}$&$3.99\times 10^{-4}$&  $\mathbf{4.22\times 10^{-4}}$ &$6.78\times 10^{-4}$\\
 $P_{12}$& $6.32\times 10^{-5}$&$1.29\times 10^{-4}$&  $3.34\times 10^{-5}$ &$4.15\times 10^{-4}$\\
\hline
\end{tabular}}
\caption{Final values of the dissociation and transition probabilities $|\langle v \ket{\chi_1(T_{\text{tot}}) } |^2$ for several bound states of H$_2^+$, for the 4 cases defined in table \ref{valnumEPH2}. The four largest values are in bold type 
and the probability which corresponds to the initial vibrational level is underlined.
}
\label{probasEPH2_tab}
\end{table}

\begin{figure}[htp]
 \centering
\includegraphics[width=0.8\linewidth]{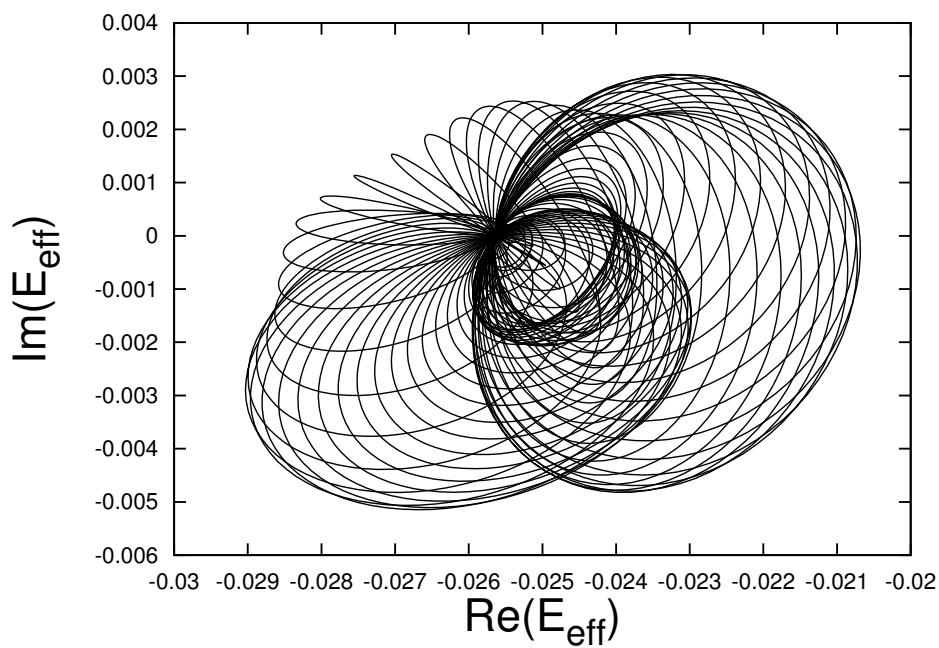}
 \caption{Indirect manifestation of the non-adiabaticity: Complex trajectory followed by 
 $E_{\text{eff}}$  (\ref{eqEeff}), for run (d), cf. table~\ref{valnumEPH2} 
  (H$_2^+$ case). Atomic units are used (1~a.u. $\simeq 27.2$ eV). 
 }
 \label{figEeff3}
\end{figure}

\begin{itemize}
 \item Case (a) : levels $v=8,7,9$ are mainly occupied at the end. Non-adiabatic exchanges seem quite strong. Final $P_{\text{diss}} \simeq 0.71 $.
 \item Case (b) : probabilities are shared between states $v=9$ and $v=8$ even if $v=9$ is predominantly occupied. The remark on non-adiabatic exchanges remains valid. 
 \item Case (c) : the initial state $v=8$ is mainly occupied at the end of the pulse. However transitions to levels $v=7,10,11$ are not negligible, $P_9$ being only the fifth larger transition. The final dissociation is estimated at $0.86$.
 \item Case (d) : there is a slight transfer from $v=9$ to $v=8$ but no clear inversion, still with non-adiabatic contaminations. 
\end{itemize} 
Most of these results disagree with the predictions of the adiabatic Floquet formalism. In terms of the occupation of the 
instantaneous Floquet eigenstates, the above results suggest that non-adiabatic exchanges happen. At least one of the hypotheses of the adiabatic approximation 
is not satisfied.

In the case of an adiabatic tracking, the presence of a sufficient distance between eigenvalues is one of the hypothesis 
of the adiabatic approximation, and non-adiabatic contaminations are probably due to the closeness of the two Floquet eigenvalues 
in the neighborhood of an EP. The total duration may also be too short. Unfortunately in the H$_2^+$ example a longer pulse would 
produce a totally dissociative process. 
Moreover the dissipation tends to depopulate the most dissipative adiabatic states. Thus even small non-adiabatic transitions can become significant and jeopardize the control, as already pointed out in \cite{uzdin,gilary2}. 
However the expected asymmetry of the flip \cite{uzdin,gilary2} is not sufficient to explain these results. 
(an adiabatic flip should be observed in half of the presented cases, and actually it is not). 



An effective eigenenergy can be calculated, which is related to the adiabatic nature of the dynamics relative to the instantaneous Floquet eigenbasis:
\begin{equation}
 E_{\text{eff}} (t) = \frac{\langle i \vert H \ket{\Psi (t) }}{\langle i \ket{\Psi(t)}}
 \label{eqEeff}
\end{equation}
where $i$ stands for the initial vibrational state. This effective energy arises from the time-dependent wave-operator formalism \cite{jolicard2003}. If the wavefunction is totally projected on a single instantaneous Floquet eigenstate $\ket{\Phi(t)}$ (as it is assumed in \cite{lefebvre}), then the above quantity takes the form
\begin{equation}
E_{\text{eff}} \simeq E_{\Phi} + i\hbar \frac{\langle i \ket{ \frac{\partial}{\partial t} \Phi (t)}}{\langle i \ket{\Phi (t)}} ,
\end{equation}
where $E_{\Phi}$ is in a sense the constant part of $E_{\text{eff}}$ and $i\hbar \frac{\langle i \ket{ \frac{\partial}{\partial t} \Phi (t)}}{\langle i \ket{\Phi(t)}}$ is a corrective term which is time-periodic. $E_{\text{eff}}$ follows a cyclic path in the complex plane. By a perturbative analyis it can be shown that an adiabatic evolution produces an elliptic path, as shown in figure 2 of \cite{jolicard2003}. Plotting $E_{\text{eff}}$ thus provides an indicator of the adiabatic nature of the dynamics.
We have calculated $E_{\text{eff}}$ for the pulse (d), as shown in figure~\ref{figEeff3}. The curve is quite complicated, with pseudo-elliptic curves which do not close on themselves. 
The most severe deformations of these curves occur when the frequency or intensity vary rapidly: A change in the orientation of the pseudo-ellipse axes is observed when the wavelength varies, while a rapid intensity variation changes the length of the axes.

\begin{figure}%
\centering
\includegraphics[width=0.8\linewidth]{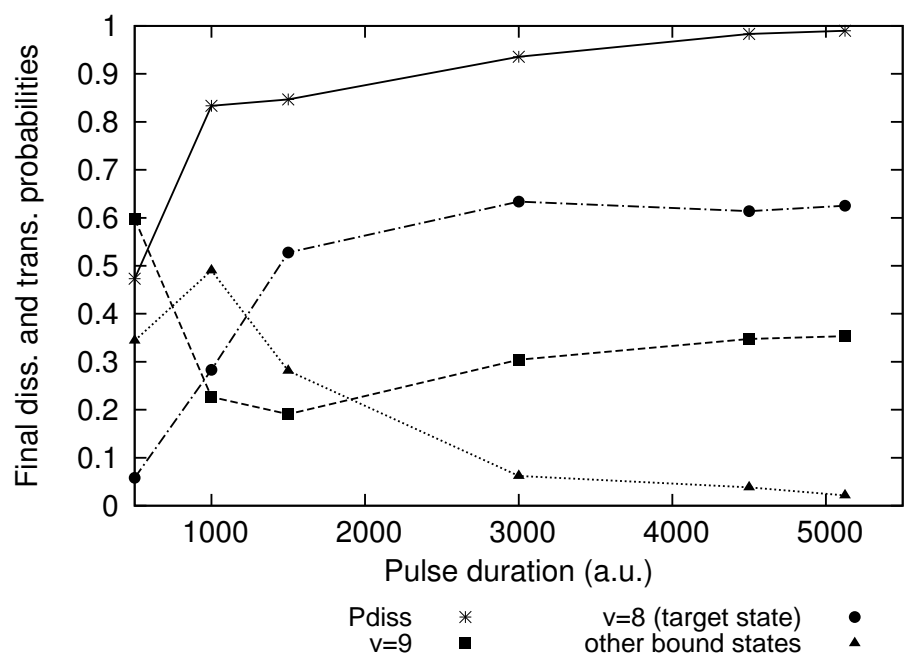}
\caption{$P_{\text{diss}}$ and fraction of the non-dissociated wavepacket in the initial state $v_i=9$ (squares), in the target state $v_t=8$ (rounds), and in all other bound state $v \neq v_i,v_t$ (triangles), calculated at the end of the pulse (d) of table \ref{valnumEPH2} (H$_2^+$ case), as a function of the pulse duration $T_{\text{tot}}$.
}%
\label{vartpsh2}%
\end{figure}

$v_i$ and $v_t$ being the initial and target bound states, we calculate the final values of $P_{v_i} /\left( \sum_{v} P_v \right)$ (fraction of the non-dissociated wavepacket surviving in initial state), $P_{v_t} / \left(\sum_v P_v \right)$ (fraction in the target state) and $\left( \sum_{v\neq v_i,v_f} P_v \right)/\left( \sum_v P_v \right)$ (other bound states).
Using these quantities, figure \ref{vartpsh2} shows what percentage of the non-dissociated wavepacket is actually flipped or not at the end of pulse, as a function of pulse duration, together with the final dissociation probability. 
These results correspond to case (d) of table \ref{valnumEPH2}. The target state population increases with increasing pulse duration but stabilises around 60 \%, while the dissociation becomes very strong and prevents us to obtain any efficient flip. 

\begin{figure}%
\centering
\includegraphics[width=0.8\linewidth]{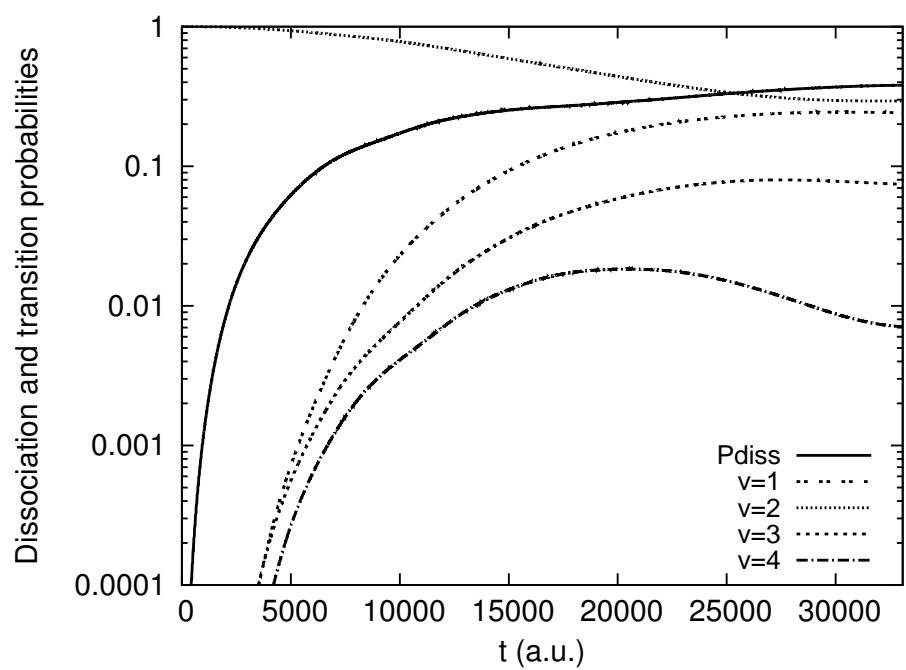}
\caption{Same as figure \ref{fig:pdissvEPH2} but for Na$_2$ and pulse (d') of table \ref{valnumEPNa2}.}%
\label{fig:pdissvna2}%
\end{figure}


Na$_2$ seemed to be a more promising case, because the photodissociation is expected to be smaller even with longer pulses. 
Final dissociation and transition probabilities for the different cases of table \ref{valnumEPNa2} are summarized in table~\ref{probasEPNa2_tab}. 
The detailed time evolution is shown in figure~\ref{fig:pdissvna2}.
These results indicate a more adiabatic behaviour. However, we do not see the expected inversions, since the wavefunction always 
ends with a maximum component on the initial vibrational state. In case (b) a partial transfer is observed from $v=3$ to $v=2$, 
even though it was level $v=4$ which was implied by the encircled EP.
The dissociation probabilities are in general larger than the values predicted using the adiabatic formalism [for instance cases (a') and (e')] 
and sometimes weaker [case (b')]. 
The five runs (a')-(e') were directly aimed at testing results of the adiabatic theory. The last one is a checking calculation with a 
larger contour in the parameter plane; it does not produce clear inversion any more.

\begin{center}
 
  \begin{table}[htp]
 \centering
{\small
\begin{tabular}{lllllll}
\hline
\hline
 Cas & (a') & (b') & (c') & (d') & (e') & (f') \\ 
\hline
 $P_{\text{diss}}$ & $0.427$ & $0.322$ & $0.467$ & $0.3805$ & $0.3438$ & $0.3922$ \\
 $P_0$ & $4.76\times 10^{-4}$ & $3.87\times 10^{-4}$ & $3.32\times 10^{-3}$ & $7.96\times 10^{-4}$ & $5.76\times 10^{-4}$ & $1.46\times 10^{-3}$ \\
 $P_1$ & $2.84\times 10^{-3}$ & $5.91\times 10^{-3}$ & $\mathbf{7.15\times 10^{-2}}$ & $\mathbf{0.2422}$ & $2.23\times 10^{-2}$ & $1.44\times 10^{-2}$ \\
 $P_2$ & $\mathbf{8.93\times 10^{-2}}$ & $\mathbf{0.2485}$ & $\mathbf{7.26\times 10^{-2}}$ & $\underline{\mathbf{0.2941}}$ & $\mathbf{9.36\times 10^{-2}}$ & $\mathbf{8.00\times 10^{-2}}$\\
 $P_3$ & $\mathbf{5.55\times 10^{-2}}$ & $\underline{\mathbf{0.3463}}$ & $ \underline{\mathbf{0.3009}}$ & $\mathbf{7.45\times 10^{-2}}$ & $\mathbf{7.13\times 10^{-2}}$ & $\mathbf{0.2157}$ \\
 $P_4$ & $\underline{\mathbf{0.3412}}$ & $\mathbf{6.59\times 10^{-2}}$ & $\mathbf{6.98\times 10^{-2}}$ & $\mathbf{7.11\times 10^{-3}}$ & $\underline{\mathbf{0.3965}}$ & $\underline{\mathbf{0.1884}}$ \\
 $P_5$ & $\mathbf{6.65\times 10^{-2}}$&$ \mathbf{9.97\times 10^{-3}}$ & $6.40\times 10^{-3}$ & $1.82\times 10^{-4}$ & $\mathbf{5.52\times 10^{-2}}$ & $\mathbf{8.11\times 10^{-2}}$\\
 $P_6$ & $1.13\times 10^{-2}$ & $3.92\times 10^{-4}$ & $9.34\times 10^{-4}$ & $3.03\times 10^{-4}$ & $7.90\times 10^{-3}$ & $3.78\times 10^{-3}$\\
 $P_7$ & $6.64\times 10^{-4}$ & $7.26\times 10^{-5}$ & $2.88\times 10^{-3}$ & $1.06\times 10^{-4}$ & $1.14\times 10^{-3}$ & $3.43\times 10^{-3}$\\
$P_8$& $1.15\times 10^{-3}$& $1.22\times 10^{-4}$&  $2.40\times 10^{-3}$ &$3.57\times 10^{-5}$&$2.38\times 10^{-3}$&$7.31\times 10^{-3}$\\
$P_9$& $1.61\times 10^{-3}$& $4.28\times 10^{-5}$&  $1.27\times 10^{-3}$ &$7.03\times 10^{-5}$&$2.37\times 10^{-3}$&$5.99\times 10^{-3}$\\
$P_{10}$& $1.28\times 10^{-3}$& $3.10\times 10^{-6}$&  $5.41\times 10^{-4}$ &$8.75\times 10^{-5}$&$1.58\times 10^{-3}$&$3.52\times 10^{-3}$\\
\hline

\end{tabular}}
\caption{Same as table \ref{probasEPH2_tab} but for the Na$_2$ model and pulses defined in table \ref{valnumEPNa2}.}
\label{probasEPNa2_tab}
\end{table}

\end{center}

\begin{figure}%
\centering
\includegraphics[width=0.8\linewidth]{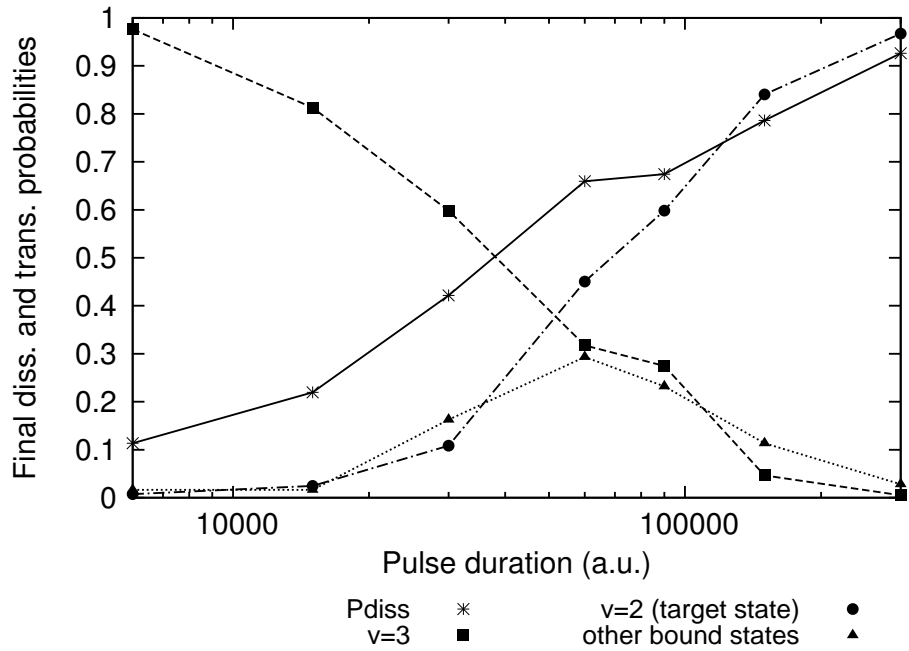}
\caption{$P_{\text{diss}}$ and fraction of the non-dissociated wavepacket in the initial state $v_i=3$ (squares), in the target state $v_t=2$ (rounds), and in all other bound state $v \neq v_i,v_t$ (triangles), calculated at the end of the pulse (c') of table \ref{valnumEPNa2} (Na$_2$ case), as a function of the pulse duration $T_{\text{tot}}$.
}%
\label{vartpsna2_1}%
\end{figure}
\begin{figure}%
\centering
\includegraphics[width=0.8\linewidth]{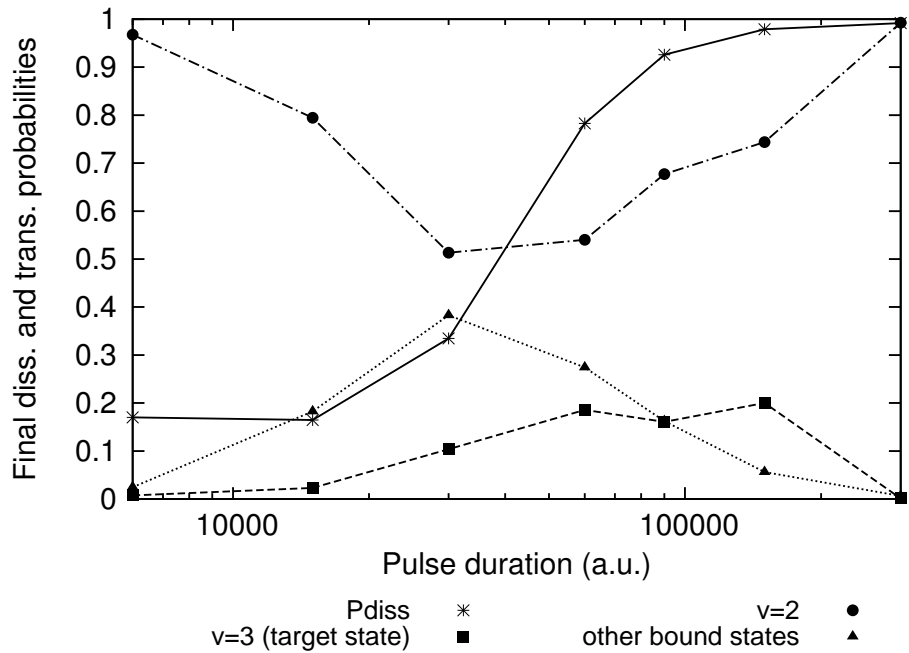}
\caption{$P_{\text{diss}}$ and fraction of the non-dissociated wavepacket in the initial state $v_i=2$ (rounds), in the target state $v_t=3$ (squares), and in all other bound state $v \neq v_i,v_t$ (triangle), calculated at the end of the pulse (d') of table \ref{valnumEPNa2}, as a function of the pulse duration $T_{\text{tot}}$.
}%
\label{vartpsna2_2}%
\end{figure}

Choosing a longer duration allows us to partially obtain the adiabatic flip from $v=3$ to $v=2$ but with a stronger dissociation than the value given by previous studies. This is shown in figure \ref{vartpsna2_1} which corresponds to case (c') of table 2 with varying pulse duration. There is no flip for a pulse duration shorter than 50000 a.u. (i.e. 1.2 ps). With $T_{\text{tot}}=150000$ a.u. (i.e. about five times the duration proposed in \cite{atabek2}), the dissociation probability becomes quite large (78\%) and 84\% of the non dissociated wavepacket ends in the target state $v=2$. With $T_{\text{tot}} = 300000$ a.u. (i.e. 7.25 ps), 96.7 \% of the 7.4 \% surviving population occupies the target state at the end of the pulse. 
The dynamics issued from initial state $v=2$ (figure \ref{vartpsna2_2}) does not produce any state interchange at any time scale, which illustrates the asymmetry of the adiabatic flip as explained in \cite{uzdin}. 

\section{Concluding remarks}

For diatomic molecules such that H$_2^+$ et Na$_2$ submitted to chirped pulses, 
wavepacket propagation calculations do not confirm previously published results from the adiabatic Floquet theory.
For the H$_2^+$ case, no controlled transfer of vibrational population is obtained and many non-adiabatic contaminations seem to jeopardize the expected control scheme. For the Na$_2$ molecule, a controlled transfer occurs only with pulse whose duration is much longer than expected from previous works, in a way that the dissociation probability becomes large. We also confirm that the interchange is only possible for only one of the two states concerned with the EP (asymmetry of the flip).
However, after these somewhat disappointing results, we are now working to improve the design of chirped laser pulses
so as to limit the non-adiabatic exchanges. 
This is undoubtedly a difficult task because the presence of long pulses, essential to create adiabaticity, simultaneously
generates totally dissociative processes. Moreover it would be necessary to leave the pure adiabaticity concept for
generalized adiabaticity, by working in the framework of the wave operator
theory \cite{reviewgeorges2}, in order to take into account the mixing of the two Floquet states near the EP.

We are 
also working on ideas of control using global Floquet eigenvectors instead of
progressive instantaneous Floquet eigenvectors. The adiabatic scheme is probably too strong of an assumption. The
control of the full transfer $\ket{i} \rightarrow \ket{f}$ implies a generalized cyclicity, i.e. a less strong constraint. 
The generalized cyclicity is the ability of the molecule issued from the subspace $S_o$ (with projector $P_o=
\ket{i}\langle i\vert +\ket{f}\langle f \vert$) to come back into this subspace at the end of the interaction. Using an extended Hilbert space
(with time being included as a new quantum variable) and the wave operator formulation, the dynamical problem
is transformed into an inverse stationary problem. We should select the adiabatic parameter evolutions
($E(t), \omega (t)$) which supress the wave-operator components outside the subspace $S_o$ at the end of the
interaction.
The control problem can 
then be formulated in terms of the cyclicity of a small numbers of global eigenvectors instead
of needing to use an adiabatic hypothesis throughout the pulse.

\section{References}

\bibliography{Leclerc_failure_adiabatic_6}

\begin{thebibliography}{10}

\bibitem{NHQM}
N.~Moiseyev.
\newblock {\em Non-Hermitian Quantum Mechanics}.
\newblock Cambridge University Press, Cambridge, 2011.

\bibitem{kato}
T.~Kato.
\newblock {\em Perturbation theory of linear operators}.
\newblock Springer, Berlin, 1966.

\bibitem{heiss2012}
W.~D. Heiss.
\newblock The physics of exceptional points.
\newblock {\em J. Phys. A: Math. Gen.}, 45:444016, 2012.

\bibitem{chap_EP_moiseyev}
N.~Moiseyev.
\newblock {\em Non-Hermitian Quantum Mechanics}, chapter~9.
\newblock Cambridge University Press, Cambridge, 2011.

\bibitem{dembowski2001}
C.~Dembowski, H.~D. Gr\"af, H.~L. Harney, A.~Heine, W.~D. Heiss, H.~Rehfeld,
  and A.~Richter.
\newblock Experimental observation of the topological structure of exceptional
  points.
\newblock {\em Phys. Rev. Lett.}, 86:787, 2001.

\bibitem{dietz}
B.~Dietz, H.~L. Harney, O.~N. Kirillov, M.~Miski-Oglu, A.~Richter, and
  F.~Sch\"afer.
\newblock Exceptional points in a microwave billiard with time-reversal
  invariance violation.
\newblock {\em Phys. Rev. Lett.}, 106:150403, 2011.

\bibitem{longhi2010}
S.~Longhi.
\newblock Optical realization of relativistic non-hermitian quantum mechanics.
\newblock {\em Phys. Rev. Lett.}, 105:013903, 2010.

\bibitem{klaiman2008}
S.~Klaiman, U.~G\"unther, and N.~Moiseyev.
\newblock Visualization of branch points in pt -symmetric waveguides.
\newblock {\em Phys. Rev. Lett.}, 101:080402, 2008.

\bibitem{lefebvre}
R.~Lefebvre, O.~Atabek, M.~\v{S}indelka, and N.~Moiseyev.
\newblock Resonance coalescence in molecular photodissociation.
\newblock {\em Phys. Rev. Lett.}, 103:123003, 2009.

\bibitem{atabek2}
O.~Atabek, R.~Lefebvre, M.~Lepers, A.~Jaouadi, O.~Dulieu, and V.~Kokoouline.
\newblock Proposal for a laser control of vibrational cooling in na$_2$ using
  resonance coalescence.
\newblock {\em Phys. Rev. Lett.}, 106:173002, 2011.

\bibitem{lefebvre2}
R.~Lefebvre, A.~Jaouadi, O.~Dulieu, and O.~Atabek.
\newblock Laser cooling of the vibrational motion of na$_2$ combining the
  effects of zero-width resonances and exceptional points.
\newblock {\em Phys. Rev. A}, 84:043428, 2011.

\bibitem{lefebvre3}
R.~Lefebvre, A.~Jaouadi, and O.~Atabek.
\newblock Clusters of exceptional points for a laser control of selective
  vibrational transfer.
\newblock {\em Chem. Phys.}, 399:111, 2012.

\bibitem{atabek3}
O.~Atabek and R.~Lefebvre.
\newblock Laser control of vibrational transfer based on exceptional points.
\newblock {\em J. Phys. Chem.}, 114:3031, 2010.

\bibitem{liertzer2012}
M.~Liertzer, L.~Ge, A.~Cerjan, A.~D. Stone, H.~E. T\"ureci, and S.~Rotter.
\newblock Pump-induced exceptional points in lasers.
\newblock {\em Phys. Rev. Lett.}, 108:173901, 2012.

\bibitem{heiss2005}
W.~D. Heiss, F.~G. Scholtz, and H.~B. Geyer.
\newblock The large n behaviour of the lipkin model and exceptional points.
\newblock {\em J. Phys. A: Math. Gen.}, 38:1843, 2005.

\bibitem{fedorov1}
M.~V. Fedorov and A.~M. Movsesian.
\newblock Field-induced effects of narrowing of photoelectron spectra and
  stabilisation of rydberg atoms.
\newblock {\em J. Phys. B: At. Mol. Opt. Phys.}, 21:L155--L158, 1988.

\bibitem{fedorov2}
M.~V. Fedorov, N.~P. Poluektov, A.~M. Popov, O.~V. Tikhonova, V.~Y. Kharin, and
  E.~A. Volkova.
\newblock Interference stabilization revisited.
\newblock {\em IEEE J. of selected topics in quantum electronics},
  18(1):42--53, 2012.

\bibitem{shirley}
J.~H. Shirley.
\newblock {\em Phys. Rev.}, 138:B979, 1965.

\bibitem{reviewguerin}
S.~Gu\'erin and H.~R. Jauslin.
\newblock Control of quantum dynamics by laser pulses: adiabatic floquet
  theory.
\newblock {\em Advances in Chem. Phys.}, 125:147, 2003.

\bibitem{CAP2}
J.~G. Muga, J.~P. Palao, B.~Navarro, and I.~L. Egusquiza.
\newblock Complex absorbing potentials.
\newblock {\em Phys. Rep.}, 395:357, 2004.

\bibitem{moiseyev1998}
N.~Moiseyev.
\newblock Quantum theory of resonances: calculating energies, widths and
  cross-sections by complex scaling.
\newblock {\em Phys. Rep.}, 302:211, 1998.

\bibitem{uzdin}
R.~Uzdin, A.~Mailybaev, and N.~Moiseyev.
\newblock On the observability and asymmetry of adiabatic state flips generated
  by exceptional points.
\newblock {\em J. Phys. A: Math. Gen.}, 44:435302, 2011.

\bibitem{gilary2}
I.~Gilary and N.~Moiseyev.
\newblock Asymmetric effect of slowly varying chirped laser pulses on the
  adiabatic state exchange of a molecule.
\newblock {\em J. Phys. B: At. Mol. Opt. Phys.}, 45:051002, 2012.

\bibitem{leclercviennot}
A.~Leclerc, D.~Viennot, and G.~Jolicard.
\newblock The role of the adiabatic phases in adiabatic population tracking for
  non-hermitian hamiltonians.
\newblock {\em J. Phys. A: Math. Gen.}, 45:415201, 2012.

\bibitem{berrycycling}
M.~V. Berry and R.~Uzdin.
\newblock Slow non-hermitian cycling: exact solutions and the stokes
  phenomenon.
\newblock {\em J. Phys. A: Math. Gen.}, 44:435303, 2011.

\bibitem{nenciurasche}
G.~Nenciu and G.~Rasche.
\newblock On the adiabatic theorem for nonself-adjoint hamiltonians.
\newblock {\em J. Phys. A: Math. Gen.}, 25:5741, 1992.

\bibitem{bunkin}
F.~V. Bunkin and I.~I. Tugov.
\newblock Multiphoton processes in homopolar diatomic molecules.
\newblock {\em Phys. Rev. A}, 8(2):601, 1973.

\bibitem{magnier}
S.~Magnier, P.~Millié, O.~Dulieu, and F.~Masnou-Seeuws.
\newblock Potential curves for the ground and excited states of the na$_2$
  molecule up to the (3s+5p) dissociation limit: Results of two different
  effective potential calculations.
\newblock {\em J. Chem. Phys.}, 98:7113, 1993.

\bibitem{fatemi}
F.~K. Fatemi, K.~M. Jones, P.~D. Lett, and E.~Tiesinga.
\newblock Ultracold ground-state molecule production in sodium.
\newblock {\em Phys. Rev. A}, 66:053401, 2002.

\bibitem{aymar2005}
M.~Aymar and O.~Dulieu.
\newblock Calculation of accurate permanent dipole moments of the lowest
  1,3$\sigma$+ states of heteronuclear alkali dimers using extended basis sets.
\newblock {\em J. Chem. Phys.}, 122:204302, 2005.

\bibitem{jaouadi}
A.~Jaouadi.
\newblock {\em Private communication}, 2012.

\bibitem{rom}
N~Rom, N.~Lipkin, and N.~Moiseyev.
\newblock Optical potentials by the complex coordinate method.
\newblock {\em Chem. Phys.}, 151:199, 1991.

\bibitem{rissmeyer}
U.~V. Riss and H.~D. Meyer.
\newblock Calculation of resonance energies and widths using the complex
  absorbing potential method.
\newblock {\em J. Phys. B}, 26:4503, 1993.

\bibitem{santra}
R.~Santra.
\newblock Why complex absorbing potentials work: A
  discrete-variable-representation perspective.
\newblock {\em Phys. Rev. A}, 74:034701, 2006.

\bibitem{CATM}
G.~Jolicard, D.~Viennot, and J.~P. Killingbeck.
\newblock Constrained adiabatic trajectory method.
\newblock {\em J. Phys. Chem. A}, 108(41):8580--8589, 2004.

\bibitem{CATM2}
A.~Leclerc, S.~Gu\'erin, G.~Jolicard, and J.~P. Killingbeck.
\newblock Quantum dynamics by the constrained adiabatic trajectory method.
\newblock {\em Phys. Rev. A}, 83:032113, 2011.

\bibitem{CATM3}
A.~Leclerc, G.~Jolicard, and J.~P. Killingbeck.
\newblock Development of a general time-dependent absorbing potential for the
  contrained adiabatic trajectory method.
\newblock {\em J. Chem. Phys}, 134:194111, 2011.

\bibitem{CATM4}
A.~Leclerc, G.~Jolicard, D.~Viennot, and J.~P. Killingbeck.
\newblock Constrained adiabatic trajectory method: A global integrator for
  explicitly time-dependent hamiltonians.
\newblock {\em J. Chem. Phys}, 136:014106, 2012.

\bibitem{jolicard2003}
G.~Jolicard, O.~Atabek, M.~L. Dubernet-Tuckey, and N.~Balakrishnan.
\newblock Nonadiabatic molecular response to short, intense laser pulses: a
  wave operator generalized floquet approach.
\newblock {\em J. Phys. B}, 36:2777, 2003.

\bibitem{reviewgeorges2}
G.~Jolicard and J.P. Killingbeck.
\newblock The bloch wave operator: generalizations and applications: ii. the
  time-dependent case.
\newblock {\em J. Phys. A}, 36(40):R411--R473, 2003.

\end{thebibliography}

\end{document}